\documentclass[superscriptaddress,amsmath,amssymb,11pt]{article}
\usepackage{graphicx,bm}
\usepackage[dvips]{color}
\usepackage{authblk}
\usepackage{amsfonts}
\usepackage[labelfont=bf]{caption}            
\newcommand{\HTO}{$\rm Ho_2Ti_2O_7$}
\newcommand{\HYTO}{Ho$_{2-x}$Y$_{x}$Ti$_2$O$_7$}
\newcommand{\DTO}{$\rm Dy_2Ti_2O_7$}
\newcommand{\RTO}{$\rm R_2Ti_2O_7$}
\newcommand{\YTO}{$\rm Y_2Ti_2O_7$}
\newcounter{firstbib}

\title{Low-temperature muon spin rotation studies of the monopole charges and currents in Y doped Ho$_2$Ti$_2$O$_7$}
\author[1]{L. J. Chang}
\author[2]{M. R. Lees}
\author[2]{G. Balakrishnan}
\author[3]{\authorcr Y. -J. Kao}
\author[4]{A. D. Hillier}
\affil[1]{Department of Physics, National Cheng Kung University,\authorcr Tainan 70101, Taiwan}
\affil[2]{Department of Physics, University of Warwick,\authorcr Coventry, CV4 7AL, United Kingdom}
\affil[3]{Department of Physics and Center of Quantum Science and Engineering, National Taiwan University, Taipei 10607, Taiwan}
\affil[4]{STFC Rutherford Appleton Laboratory, ISIS Facility,\authorcr Didcot OX11 0QX, United Kingdom}

\begin{document}
\date{}
\maketitle

\begin{abstract}
In the ground state of \HTO\ spin ice, the disorder of the magnetic moments follows the same rules as the proton disorder in water ice. Excitations take the form of magnetic monopoles that interact via a magnetic Coulomb interaction. Muon spin rotation has been used  to probe the low-temperature magnetic behaviour in single crystal \HYTO\ ($x=0$, 0.1, 1, 1.6 and 2).  At very low temperatures, a linear field dependence for the relaxation rate of the muon precession $\lambda(B)$, that in some previous experiments on $\rm Dy_2Ti_2O_7$ spin ice has been associated with monopole currents, is observed in samples with $x=0$, and 0.1. A signal from the magnetic fields penetrating into the silver sample plate due to the magnetization of the crystals is observed for all the samples containing Ho allowing us to study the unusual magnetic dynamics of Y doped spin ice.
\end{abstract}

\newpage
In the spin ice materials \RTO\ (R = Ho, Dy)~\cite{Ramirez,Harris,Bramwell1} a large ($\sim\!10\mu_B$) magnetic moment on the R$^{3+}$ ions giving a strong, but at low temperature almost completely screened dipole-dipole interaction, together with a local Ising-like anisotropy leads to an effective nearest-neighbour frustrated ferromagnetic interaction between the magnetic moments. The organizing principles of the magnetic ground-state in spin ice, or ``ice rules", require that two R$^{3+}$ spins should point in and two out of each elementary tetrahedron in the \RTO\ pyrochlore lattice~\cite{Harris,Bramwell2,Hertog,Yavorskii,Isakov}. Excitations above the ground state manifold, which locally violate the ice rules, can be viewed as magnetic monopoles of opposite ``magnetic charge" connected by Dirac strings~\cite{Castelnovo,Ryzhkin,Jaubert}. Evidence of magnetic monopoles in spin ice has recently been observed in several experiments~\cite{Fennell,Morris,Kadowaki}. 

Given the existence of magnetic monopoles, it is logical to consider the nature of the magnetic charges and any associated currents or ``magnetricity". Bramwell \textit{et al}. used transverse-field muon spin-rotation (TF-$\mu$SR) to investigate the magnitude and dynamics of the magnetic charge in \DTO\ spin ice~\cite{Bramwell3}. In these experiments the equivalence of electricity and magnetism proposed in Ref.~\cite{Castelnovo} was assumed and Onsager's theory~\cite{Onsager}, which describes the nonlinear increase with applied field in the dissociation constant of a weak electrolyte (second Wien effect), was applied to the problem of spin ice. It was argued that in spin ice, if the magnetic field $B$ is changed, the relaxation of the magnetic moment $\nu_{\mu}$ occurs at the same rate as that of the monopole density and so in the weak field limit, $\nu_{\mu}(B)/\nu_{\mu}(0)=\kappa(B)/\kappa(0)=1+b/2$, where $\kappa$ is the magnetic conductivity and $b=\mu_0Q^3B/8\pi k_B^2T^2$ with a magnetic charge $Q$~\cite{Bramwell3}. At low temperature, the fluctuating local fields lead to a de-phasing of the muon precession and an exponential decay in the oscillatory muon polarization as a function of time $t$ 
\begin{equation}
A(t)=A_0\cos(2\pi \upsilon t)\exp(-\lambda t),
\label{Exponential decay}
\end{equation}
where $A_0$ is the initial muon asymmetry, $\upsilon=\gamma_{\mu}B/2\pi$ is the frequency of the oscillations, and  $\gamma_{\mu}$ is the gyromagnetic ratio.
With $\nu_{\mu}(B)/\nu_{\mu}(0)=\lambda(B)/\lambda(0)$ one can directly infer the magnetic monopole charge. These measurements have proven intriguing and controversial. Dunsiger \textit{et al}. ~\cite{Dunsiger} contend that the TF-$\mu$SR data never takes a form where $\lambda\propto\nu$ (see however~\cite{Bramwell6}). It has also been suggested that the magnetic field at any muon implantation site in \DTO\ is likely to take a range of values up to 0.5~T~\cite{Dunsiger,Lago,Blundell}. If this is the case it is difficult to understand how the  fields of 1-2~mT used in Ref.~\cite{Bramwell3} could lead to a precession signal. Both Dunsiger \textit{et al}.~\cite{Dunsiger} and later Blundell~\cite{Blundell} have suggested that the signals seen in the $\mu$SR data in Ref.~\cite{Bramwell3} originate from outside the sample. In their reply to this suggestion, Bramwell \textit{et al}.~\cite{Bramwell5} acknowledged that their experiments exploited both muons implanted in the sample (interior muons) and muons decaying outside the sample (exterior muons), with the aim of separating near and far field contributions to the signal. They went on to note that the signal at higher temperatures is dominated by muons implanted in the silver backing plate. This possibility was not discussed in their original paper~\cite{Bramwell3}. Nevertheless, they continued to insist that the signal at low temperature ($0.4>T>0.07$~K) cannot be explained by exterior muons and that the Wien effect signal originates from muons within the sample or muons sufficiently close to the surface of the sample so as to probe the monopolar far field.

\section*{Results}

\begin{figure}[t]%-----------FIG1--------------------------
\begin{center}
\includegraphics[width=0.7\columnwidth]{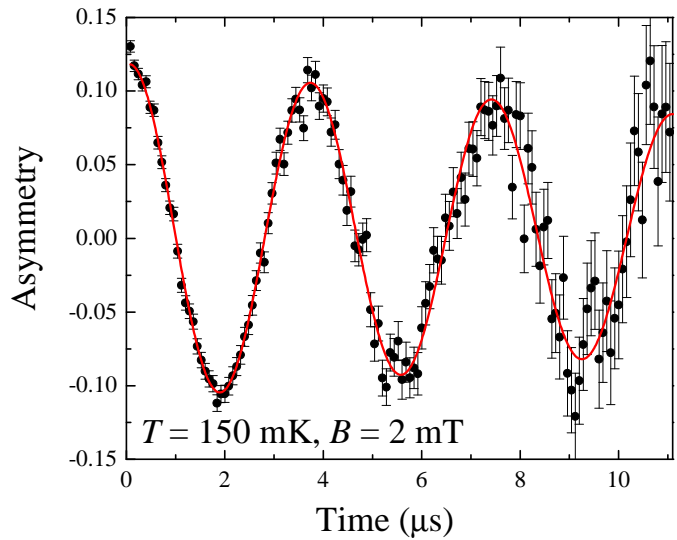}
\caption{\label{Fig1}\textbf{TF-$\boldsymbol{\mu}$SR time spectrum collected at 150~mK in a field of 2~mT for a pure Ho$\boldsymbol{_{2}}$Ti$\boldsymbol{_2}$O$\boldsymbol{_7}$ sample.} These results are representative of the data collected during this study.}
\end{center}
\end{figure}

Fig.~\ref{Fig1} shows a TF-$\mu$SR time spectrum  collected at 150~mK in a field of 2~mT for a pure \HTO\ sample. This curve is representative of the data collected during this study. A rapid loss in asymmetry from an initial value of $\sim0.22$ occurs outside the time window of the MuSR spectrometer~\cite{Bramwell3,Lago,Blundell}. The slowly relaxing component of the data were fit using Eq.~\ref{Exponential decay}. 

\begin{figure}[tb]%-----------FIG2---------------
\begin{center}
\includegraphics[width=0.7\columnwidth]{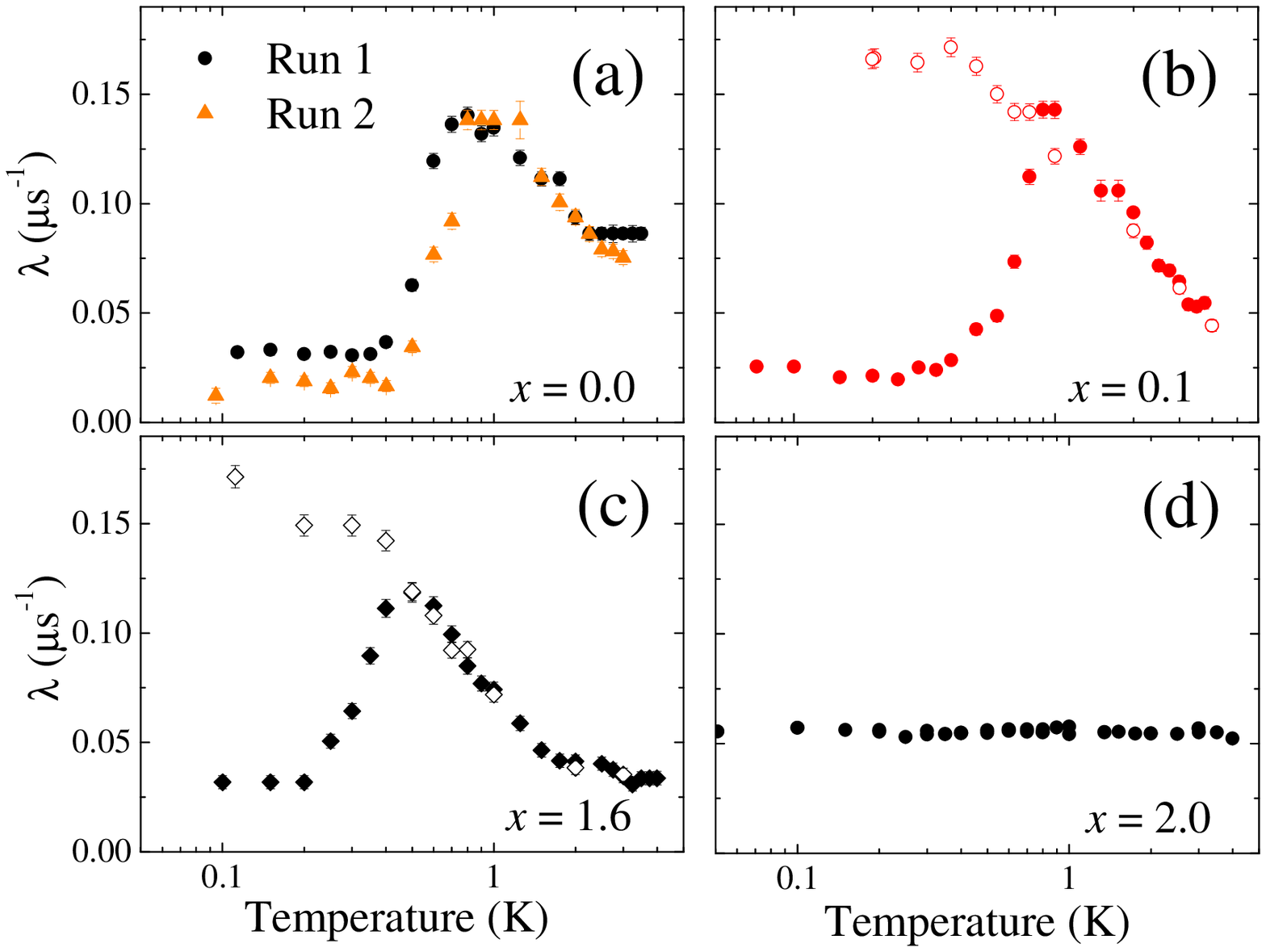}
\caption{\label{Fig2} \textbf{Temperature dependence of the muon relaxation rate $\boldsymbol{\lambda(T)}$ extracted the fits to the TF-$\boldsymbol{\mu}$SR time spectra collected in 2~mT for samples of Ho$\boldsymbol{_{2-x}}$Y$\boldsymbol{_{x}}$Ti$\boldsymbol{_2}$O$\boldsymbol{_7}$ with $\boldsymbol{x=0}$, 0.1, 1.6 and 2.0.} The closed symbols show the zero-field-cooled warming data and the open symbols show the field-cooled cooling data.}
\end{center}
\end{figure}

Fig.~\ref{Fig2} shows the temperature dependence of the muon relaxation rate $\lambda(T)$ for \HYTO\ extracted from fits to $\mu$SR time data collected in 2~mT, (see methods and \cite{SuppNote}). For all the samples containing Ho, a nearly $T$ independent $\lambda(T)$ is observed at low-temperature. As the temperature is raised there is a rapid increase in $\lambda(T)$ at some crossover temperature $T_{CR}$.  This  $T_{CR}$ increases from $\sim0.4$~K for the crystals with $x=1.6$ and 1.0 (data not shown) to 0.5~K for the samples with  $x=0.1$ and 0.0. Above $T_{CR}$ the relaxation rate decreases with increasing temperature and has a similar $T$ dependence for all four samples containing Ho that were studied. For two samples ($x=0.1$ and 1.6) we also collected field-cooled-cooling data. In both cases a divergence between the zero-field-cooled warming (ZFCW) and the field-cooled cooling (FCC) curves appears at $T_{CR}$. For pure \YTO\ a temperature independent relaxation rate is measured for the whole temperature range (0.05 to 5~K) studied. 

In order to better understand the origins of these signals we have also collected relaxation data as a function of temperature in 2~mT for the pure~\HTO\ sample discussed above, covered with a silver foil 0.25~mm thick. This thickness of foil is expected to stop all the muons before they reach the sample. Muons implanted in silver have a negligible relaxation and so any relaxation must result from a combination of the externally applied field and/or field lines originating from the sample penetrating into the silver. The $\lambda(T)$ curve obtained in this way (see ~\cite{SuppNote}) is very similar to the signal from the pure \HTO\ shown in Fig.~\ref{Fig2}a and demonstrates that at least some of the signal come from fields within the silver, but that these fields are the result of the magnetic properties of the sample~\cite{SuppNote}.

\begin{figure}[tb]%-----------FIG3---------------
\begin{center}
\includegraphics[width=0.6\columnwidth]{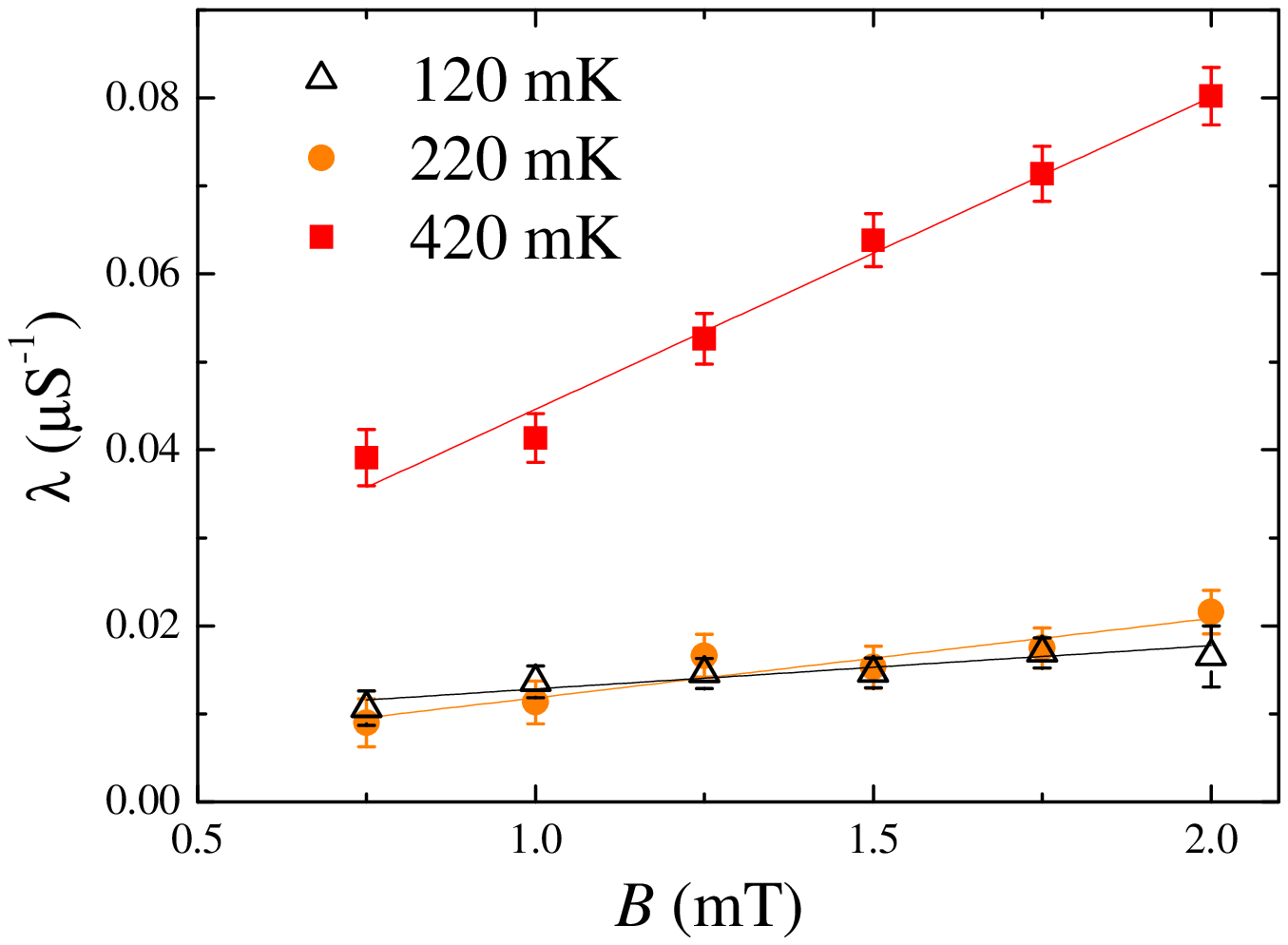}
\caption{\label{Fig3}\textbf{Magnetic field dependence of the muon relaxation rate $\boldsymbol{\lambda(B)}$ for pure Ho$\boldsymbol{_{2}}$Ti$\boldsymbol{_2}$O$\boldsymbol{_7}$ at three different temperatures.} The values for $m=(d\lambda(B)/dB)/\lambda_0$ and the effective magnetic charge $Q_{\rm{eff}}$ shown in figure~\ref{Fig4} have been obtained from the straight line fits to the data.}
\end{center}
\end{figure}

\begin{figure}[tb]%-----------FIG4---------------
\begin{center}
\includegraphics[width=0.7\columnwidth]{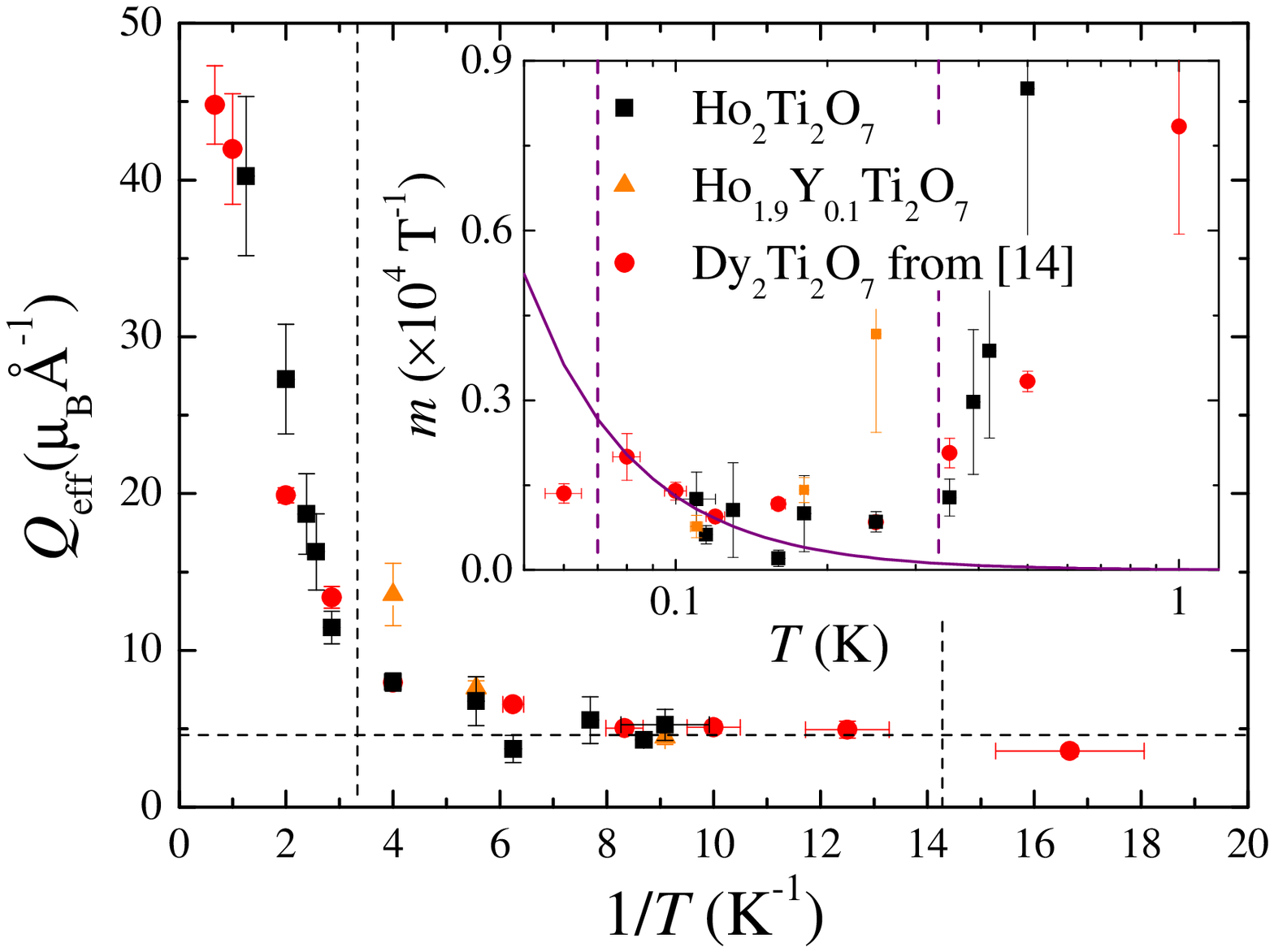}
\caption{\label{Fig4}\textbf{$\boldsymbol{Q_{\rm{eff}}}$ versus $\boldsymbol{1/T}$ for samples of Ho$\boldsymbol{_{2-x}}$Y$\boldsymbol{_{x}}$Ti$\boldsymbol{_2}$O$\boldsymbol{_7}$ with $\boldsymbol{x=0}$ and 0.1.} The vertical dashed lines indicate the high and low temperature limits between which the Onsager theory is expected to be valid~\cite{Bramwell3} and the horizontal line marks the value for $Q_{\rm{eff}}=4.6~\mu_B$\AA$^{-1}$~\cite{Castelnovo}. The inset shows $m(T)$ for the same data; the solid line shows $m=Q_{\rm{eff}}^3/T^{2}$ with $Q_{\rm{eff}}=5~\mu_B$\AA$^{-1}$. Also shown in both plots are the data of Bramwell \textit{et al}. from Ref.~\cite{Bramwell3}.}
\end{center}
\end{figure}

As a next step we then investigated the magnetic field dependence of muon relaxation rate. Fig.~\ref{Fig3} shows $\lambda(B)$ for a sample with $x=0$ at selected temperatures. Studies were also made for samples with $x=0.1$, 1, 1.6 and 2. Following Bramwell \textit{et al}., linear fits to the $\lambda(B)$ data were made at each temperature.  Using the gradient and intercept extracted from each fit, the effective magnetic charge $Q_{\rm{eff}}$ was obtained from $Q_{\rm{eff}}=2.1223m^{1/3}T^{2/3}$, where $m=(d\lambda(B)/dB)/\lambda_0$~\cite{Bramwell3}. For samples with $x=0$ and 0.1 the resulting values of $Q_{\rm{eff}}$ range from 4.5 to $7.5~\mu_B$\AA$^{-1}$ in the temperature regime in which Onsager's theory is expected to be valid, but increase rapidly as the temperatures increase outside this range (see Fig.~\ref{Fig4}). 

At high temperature, a linear field dependence for $\lambda(B)$ is also observed for the two samples with a much higher yttrium doping ($x=1$ and 1.6) but the calculated $Q_{\rm{eff}}$ is always greater that $\sim10~\mu_B$\AA$^{-1}$. For $x=1$ and 1.6 in the low-temperature regime $T < T_{CR}$ there is no systematic linear field dependence in  $\lambda(B)$ and no signal that can be associated with magnetricity.

We have also looked for a linear magnetic field dependence in $\lambda(B)$ for the pure \HTO\ sample covered in a thick (0.25~mm) silver foil. At higher temperatures $T > T_{CR}$ we observed a linear behaviour leading to a large $Q_{\rm{eff}}$ (i.e. $Q_{\rm{eff}}> 10~\mu_B$\AA$^{-1}$), but at low temperatures $T < T_{CR}$ we found no signature of magnetricity and could not obtain reliable linear fits to the $\lambda(B)$ data or physically acceptable values for $Q_{\rm{eff}}$. \

\section*{Discussion}

We can draw a number of important conclusions from our work. Our results indicate that at higher temperatures, as suggested previously~\cite{Dunsiger,Blundell,Bramwell5}, the dominant contribution to the $\lambda(T)$ signal arises from stray fields from the magnetized spin ice that penetrate into the silver sample plate. The observation of a signal in a sample covered with thick Ag foil adds weight to this hypothesis. The sample coverage of the Ag backing plates used in our experiments was always approximately 50\%. It will be interesting to explore how this signal changes as this coverage is varied. It may also be important to consider the ratio between the surface area and the volume of the spin ice in these and other experiments. Differences between the bulk and surface conductivity of water ice are well documented~\cite{PetrenkoWhitworth} and it is likely that analogous processes operate in spin ice. In reply to the comments on their work, however, Bramwell \textit{et al}.~\cite{Bramwell5} make the point that a signal from muons implanted in the sample plate may not negate the important findings of their study. Our data are consistent with the suggestion made in Ref.~\cite{Bramwell5} that the Wien effect signal may arise from inside the sample or from within the Ag sample plate but at distances very close to the spin ice sample surface. We will return to this point later. First we note that the $\lambda(T)$ curve for pure \HTO\ follows closely the form expected for the magnetization of pure spin ice~\cite{Snyder2} supporting the view that $\lambda(T)$  reflects the magnetization in all the samples studied. This then raises an interesting question concerning the low-temperature magnetic dynamics of spin ice.

Recently there have been a number of experimental reports on the magnetic dynamics of spin ice (see for example~\cite{Giblin,Slobinsky,Yaraskavitch, Erfanifam, Petrenko}).  In addition to the discussion of magnetic monopoles and the Wien effect~\cite{Castelnovo, Ryzhkin, Jaubert, Bramwell3} authors have also considered the effects of thermal quenching~\cite{Castelnovo2}. A key component of the current theories of  spin ice, is that the magnetic response at low temperatures and small applied fields is limited to monopole motion. So as the monopole density decreases the characteristic time scales become longer. This view has recently been called into question following new low-temperature AC susceptibility measurements that exhibit an activated behaviour with energy barriers that are inconsistent with the present understanding of monopoles in spin ice~\cite{Yaraskavitch, Matsuhira, Quilliam}. Our results for the $x=1.6$ material, showing the survival of ZFCW-FCC splitting in a sample with only 15\% Ho add a further twist to this puzzle. Given the large number of non-magnetic ``defects" on the corners of many of the  tetrahedra in this diluted material, it is not easy to attribute the slow relaxation to a low monopole density. At such low concentrations of magnetic ions even the concepts of a spin ice and monopoles are questionable.

It is conceivable that single ion physics plays a more important role in the behaviour of the diluted materials. Our diffuse neutron-scattering studies of single-crystal \HYTO\ showed that at low temperature the scattering patterns are characteristic of a dipolar spin ice and appear to be unaffected by Y doping up to at least $x=1.0$~\cite{Chang}. One possible scenario is that effects, such as distortions in the local environment due to the variation in the size of the Ho$^{3+}$/Y$^{3+}$ ions~\cite{Snyder1}, produce energy barriers at low-$T$ that exceed the cost of an isolated monopole. The slow dynamics and the ZFCW-FCC hysteresis at low temperatures would thus cross over from a regime where this behaviour is attributed to low monopole density to a regime where it is due to exceedingly slow single ion physics. Alternatively, the long-range nature of the dipolar interactions may give rise to collective effects beyond the monopole description which introduce new energy barriers to spin flipping at very low temperatures that occur in both undiluted and diluted systems. The same qualitative form for the $\lambda(T)$ data for samples with $x=0.1$ and 1.6 indicate that additional ingredients may be required to explain the low $T$ behaviour in spin ice and that further studies on diluted samples are needed to fully understand the role played by factors such as impurities, dislocations, and surface effects on the low-temperature dynamics of spin ice. 

Returning to the question of magnetricity in spin ice we note that in our $\mu$SR data the low-temperature signal that has previously been interpreted as a signature of magnetricity is seen in the $x=0$ and 0.1 samples and is not observed in the more dilute \HYTO\ materials. Within the $T$ range indicated by the dashed lines in Fig.~\ref{Fig4}, where the theory presented by Bramwell \textit{et al}. is expected to be valid, the value of $Q_{\rm{eff}}$ agrees with expectations. Following Blundell~\cite{Blundell} we also plot $m$ versus $T$. We see that the expected $m\propto T^{-2}$ only holds for the same narrow $T$ range. Our experiments, including two separate runs on pure \HTO\ carried out three months apart, demonstrate the reproducibility of the data (see Fig.~\ref{Fig2}a). A realignment of the \HTO\ disks between runs also shows that the results are not particularly sensitive to the exact details of the sample geometry. Our results for the samples with a higher Y content and with the thick Ag foil demonstrate that the behaviour cannot be attributed to instrumental effects. The samples were made at Warwick~\cite{Balakrishnan} and are Ho rather than Dy based pyrochlores, eliminating the possibility of material specific results. 

In summary transverse-field $\mu$SR experiments on \HYTO, including measurements on non-magnetic \YTO\ and a sample of \HTO\ covered in thick silver foil, suggest that the majority signal in the $\lambda(T)$ response comes from stray fields due the sample magnetization penetrating into the silver sample plate~\cite{Dunsiger, Blundell}. The results for \HTO\ are comparable with those observed for \DTO. The low-temperature ($T< T_{CR}$) linear field dependence in $\lambda(B)$ is only observed in samples with $x= 0$ and 0.1. In this low-temperature regime the value of $Q_{\rm{eff}}$ agrees quantitatively with the theory presented in Ref.~\cite{Bramwell3}. The low-temperature hysteresis in $\lambda(T)$ for the magnetically dilute material ($x= 1.6$) appears inconsistent with the current understanding of monopoles in spin ice.  

\section*{Methods}
Single crystals of \HYTO\ ($x=0$, 0.1, 1, 1.6 and 2) were grown in an image furnace using the floating zone technique~\cite{Balakrishnan}. The single crystal disks were glued on to a silver plate and covered with a thin (0.01~mm) sheet of silver foil to improve thermal conductivity. The plate was then attached to the cold stage of an Oxford Instruments $^3$He/$^4$He dilution refrigerator. Transverse-field muon spin-rotation experiments were performed using the MuSR spectrometer at the ISIS pulsed muon facility, Rutherford Appleton Laboratory, UK. The magnetic field was applied along the [001] direction, perpendicular to the initial direction of the muon spin polarization which was along a [110] axis. Measurements were carried out as a function of applied field at fixed temperature and as a function of temperature in a fixed magnetic field. See~\cite{SuppNote} for full details of the measurement protocols.

\

\section*{Acknowledgements}

This work was supported by the EPSRC, United Kingdom (EP/I007210/1) and the National Science Council, Taiwan (grant no. NSC 101-2112-M-006-010-MY3). Some of the equipment used in this research was obtained through the Science City Advanced Materials project: Creating and Characterizing Next Generation Advanced Materials project, with support from Advantage West Midlands (AWM) and part funded by the European Regional Development Fund (ERDF). We would like to thank Stephen Blundell, Steve Bramwell, Claudio Castelnovo, and Sean Giblin for useful discussions.

\section*{Author contributions}

L.J.C and M.R.L. conceived of the project. G.B. prepared the samples. A.D.H, L.J.C, and M.R.L. planned and carried out the experiments. M.R.L, A.D.H,  Y.J.K, and L.J.C helped to analyse the data, draft the paper, and prepare the figures. All the authors reviewed the manuscript.

\section*{Additional information}

Competing financial interests: The authors declare no competing financial interests.

\newpage

\section*{Supplementary Information}
\subsection*{Sample Preparation}
\begin{figure}[th!]%-----------FIG1--------------------------
\begin{center}
\includegraphics[width=0.5\columnwidth]{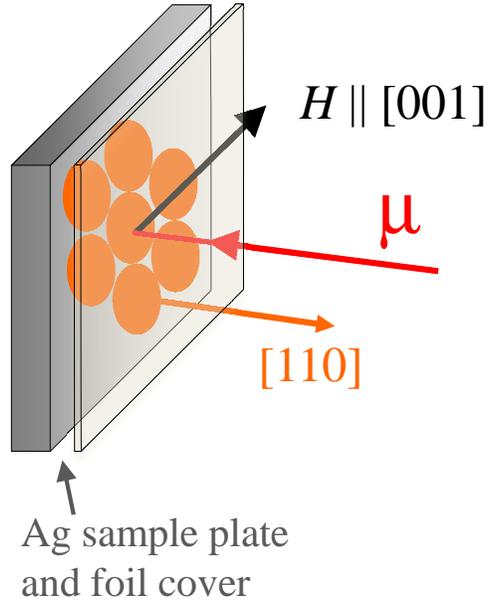}
\caption{\label{Fig1_Suppl} \textbf{Schematic of the experimental sample geometry used for the transverse-field $\boldsymbol{\mu}$SR measurements.} The single crystal disks were glued to a silver sample plate with GE varnish. The transverse field (TF) $H$ was applied along a [001] direction perpendicular to the initial direction of the muon spin polarization which was along a [110] axis.}
\end{center}
\end{figure}

Single crystals of \HYTO\ ($x=0$, 0.1, 1, 1.6 and 2) were grown in an image furnace using the floating zone technique~\cite{Balakrishnan1}. The cylindrical crystals were cut into circular disks $\sim\!6$~mm in diameter and $\sim\!1$~mm thick. These disks were oriented using the Laue x-ray diffraction technique and then glued, using GE varnish, in a circular pattern on to a silver sample plate as shown in Fig.~\ref{Fig1_Suppl}. The samples were covered with a thin (0.01~mm) sheet of silver foil to improve thermal conductivity and mounted on the cold stage of an Oxford Instruments $^3$He/$^4$He dilution refrigerator.
\newline

\subsection*{$\mathbf{\mu}$SR Experiments}

Muon spin rotation ($\mu$SR) experiments were performed using the MuSR spectrometer at the ISIS pulsed muon facility, Rutherford Appleton Laboratory, United Kingdom.

For the magnetic field sweeps at fixed temperature the samples were zero-field cooled to a temperature well below the eventual measuring temperature, thermalised, and then slowly warmed in zero field to the required measuring temperature. A transverse external field was then applied. For these measurements each field point took approximately 15 minutes to collect. At the end of each field sweep the magnetic field was reduced to zero and the sample warmed to 4~K. Note, during zero-field cooling, the stray fields at the sample position were cancelled to less than 3~$\mu$T by three pairs of coils forming an active compensation system. 

The temperature sweeps were always made following a field sweep measurement at base temperature. This means that in practice the samples were zero-field-cooled to the base temperature of the cryostat, thermalised and a field of 2 mT was applied in steps of 0.25~mT over a period of at least two hours. Data were then collected in zero-field-cooled warming (ZFCW) mode by warming the sample to each measuring temperature up to maximum of 4~K  and then in field-cooled cooling (FCC) mode on subsequent cooling to each measuring temperature. Each point in these temperature scans took around 15 minutes to collect. Due to the large low-temperature hyperfine contribution to the specific heat for the samples containing holmium, the effective base temperature of the dilution refrigerator for these samples was limited to 100~mK. 

\section*{Ho$_2$Ti$_2$O$_7$ covered in thick silver foil}

\begin{figure}[t]%-----------FIG2--------------------------
\begin{center}
\includegraphics[width=0.7\columnwidth]{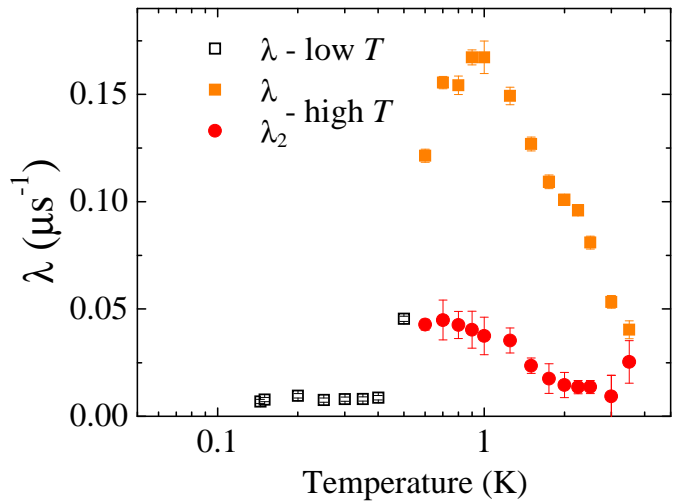}
\caption{\label{Fig2_Suppl} \textbf{Temperature dependence of the muon relaxation rate $\boldsymbol{\lambda(T)}$ extracted from fits to the TF-$\boldsymbol{\mu}$SR time spectra collected in 2~mT during ZFCW for a sample of Ho$\mathbf{_{2}}$Ti$\mathbf{_{2}}$O$\mathbf{_{7}}$ covered with 0.25 mm thick silver foil}. At low temperature a fit using Equation~\ref{Exponential decay} gives $\lambda$. At higher temperature a two component fit using Equation~\ref{Two Exponential decay} gives $\lambda$ and $\lambda_2$.}
\end{center}
\end{figure}

Muon spin rotation spectra for a sample of pure~\HTO\ covered with a silver foil 0.25~mm thick  were collected at fixed temperature in 2~mT. The temperature dependence of the muon relaxation rate $\lambda(T)$ extracted from fits to this data are shown in Fig.~\ref{Fig2_Suppl}.

For the low-temperature data ($T<T_{CR}$) the data were fit using
\begin{equation}
A(t)=A_0\cos(2\pi \upsilon t)\exp(-\lambda t),
\label{Exponential decay}
\end{equation}
where $A_0$ is the initial muon asymmetry, $\upsilon=\gamma_{\mu}B/2\pi$ is the frequency of the oscillations, and  $\gamma_{\mu}$ is the gyromagnetic ratio.

In order to obtain satisfactory fits to the data above $T_{CR}$ the modified expression. 
\begin{equation}
A(t)=A_0\cos(2\pi \upsilon t)\exp(-\lambda t)+ A_2\exp(-\lambda_2 t)
\label{Two Exponential decay}
\end{equation}
was used. The additional $A_2\exp(-\lambda_2 t)$ term is required to take account of the larger range stray fields within the thick silver foil. This is because the muon facility at ISIS has a significant momentum bite and so the implantation distance for the lower energy muons will be less than those with a higher energy.

\end{document}